# Subjective–objective policy making approach: Coupling of resident-values multiple regression analysis with value-indices, multi-agent-based simulation

Misa Owa[1*], Junichi Miyakoshi[1], Takeshi Kato[2]

[1] Hitachi Kyoto University Laboratory, Center for Exploratory Research, Research & Development Group, Hitachi Ltd., Kyoto, Japan

[2] Hitachi Kyoto University Laboratory, Open Innovation Institute, Kyoto University, Kyoto, Japan

\* Corresponding author

E-mail: misa.owa.jv@hitachi.com

Author contribution statement

# Abstract

Given the concerns around the existing subjective and objective policy evaluation approaches, this study proposes a new combined subjective–objective policy evaluation approach to choose better policy that reflects the will of citizens and is backed up by objective facts. Subjective approaches, such as the Life Satisfaction Approach and the Contingent Valuation Method, convert subjectivity into economic value, raising the question whether a higher economic value really accords with what citizens want. Objective policy evaluation approaches, such as Evidence Based Policy Making (EBPM) and Multi-Agent-Based Simulation (MABS), do not take subjectivity into account, making it difficult to choose from diverse and pluralistic candidate policies. The proposed approach establishes a subjective target function based on a multiple regression analysis of the results of a residents questionnaire survey, and uses MABS to calculate the objective evaluation indices for a number of candidate policies. Next, a new subjective–objective coupling target function, combining the explanatory variables of the subjective target function with objective evaluation indices, is set up, optimized to select the preferred policies from numerous candidates. To evaluate this approach, we conducted a verification of renewable energy introduction policies at Takaharu Town in Miyazaki Prefecture, Japan. The results show a good potential for using a new subjective-objective coupling target function to select policies consistent with the residents values for well-being from 20,000 policy candidates for social, ecological, and economic values obtained in MABS. The new approach is not just limited to multiple regression analysis and MABS but can also be developed, for example, into a combination of subjective approaches such as social impact assessment and objective ones such as EBPM. Using the new approach to compare several policies enables concrete expression of the will of stakeholders with diverse values, and contributes to constructive discussions and consensus-building.

# Introduction

The Sustainable Development Goals (SDGs) [1], adopted by the United Nations in 2015 are targets that all countries are working to achieve. Specific national initiatives include the recent Clean Energy Revolution [2], a 2021 US climate-change initiative, and Finland and Denmark, countries with high levels of SDG achievement, are working on Commitment 2050 [3] and UN 17 Village [4]. Japan's Society 5.0 [5] focuses on the importance of achieving a sustainable society. Thus, these initiatives deal with a number of social issues related to poverty, gender, food, and energy, that are increasingly recognized as common global challenges.

However, the national or local government policies introduced to deal with such issues occasionally do not take into account the participation of the citizens, a key stakeholder, or their interests, leading to opposition and protests. One reason for this is the influence of the differences in values between the stakeholders who introduce the policies and those who are affected by them. For example, in Portland, USA, in the early 1970s, local residents opposed the construction of a federally subsidized highway as part of a nation-wide policy implementation, on the grounds of environmental pollution. The city cancelled the construction of a highway in favor of a tramway linking the city and suburbs [6]. In addition, the city development office started considering residents views, for example, by building a park on the proposed highway site.

Communities of all sizes, be they national, regional or corporate, face such social challenges. Taking the regional as an example, the main value indices for citizens are their daily life, household finances and family future, while for local government the main value indices are local population, finances, and infrastructure development. Thus, there are multidimensional value indices in the region and different stakeholders have different values. Therefore, it is very important to make policy decisions that all stakeholders with different values can agree upon. In recent years, various efforts have been made to build consensus [7] [8] [9] [10].

There are two types of policy evaluation indices that various stakeholders use. One is people's subjective values, such as "like/dislike," which can be obtained through questionnaires, among others. The other is objective value indices, such as "economic strength," which can be obtained through statistical data or numerical simulations. Both value indices are important; it is not sufficient to consider only one.

There are several approaches to evaluating policies using subjective value indices, such as the Life Satisfaction Approach (LSA) [11], the Contingent Valuation Method (CVM) [12] and the Hedonic Price Method (HPM) [13], which convert non-market values into monetary values and estimate the monetary effects of policies. For example, CVM is mainly used as a method of valuing the natural environment. Specifically, it puts a monetary value on the environment by asking, for example, in a questionnaire, how much people are willing to pay to protect the environment. However, whether the monetary values for the policies are in line with subjective values, such as whether the residents are happy or satisfied with these values is unknown.

By contrast, Evidence Based Policy Making (EBPM) [14] and Multi-Agent-

Based Simulation (MABS) [15] evaluate policies using objective value indices. These approaches do not take into account the subjective values of the residents. EBPM was originally developed in the medical field and uses appropriate data to analyze and make the right decisions, rather than making decisions based on intuition and experience alone. MABS can calculate multidimensional evaluation indices and many other policies through numerical simulation. However, the output of the simulation results is too many for the stakeholders to grasp all of them, which creates a new problem: they cannot decide which result to select.

Thus, it is difficult to evaluate policies using only subjective or objective approaches. To this end, Lindsey [16] also discusses the importance of continuity in integrating the subjective and objective approaches by conducting a comparative study of each approach. This study argues that policies that will enable a region to become a sustainable, self-reliant society and enable its residents to lead happy and fulfilling lives are important. In addition, a study of Australian fooball players has been conducted to determine whether game performance indices can be explained by players' subjective ratings [17]. The field of well-being research has also studied the relationship between subjective and objective well-being [18], and these two studies also show the importance of not only subjective and objective assessments, but also the relationship between the two. Therefore, it proposes a new method of policy decision-making based not only on the objective values of the various stakeholders in the region, but also on their subjective values. Specifically, the first step is to conduct a questionnaire survey of residents as in the LSA flow, and establish response variables from the survey results. Multiple regression equations are formulated using other survey items that are highly correlated with the set response variables explanatory variables. Next, as a new

subjective–objective policy-making approach (SOPMA), a new subjective–objective coupling target function is established by substituting the objective evaluation indices of the policy simulation, obtained using MABS, into the explanatory variables for the multiple regression equation obtained earlier. Optimization with respect to this coupling target function then enables the selection of policies the most in line with subjective values from among a large number of objective policy alternatives.

To validate the SOPMA approach, we focused on the energy used in the region here. Of the energy used in the region, electricity, in particular, is generated and supplied by a major power company located outside the region, which pays for electricity outside the region. Therefore, in order to make the region sustainable and self-reliant, the policy consideration will focus on energy local production for local consumption, which can increase the circulation rate of the local economy by generating and supplying power within the region from natural sources.

The proposed SOPMA is novel in two aspects. The first is that policies can be assessed directly using a coupling target function, rather than by converting them into monetary values as in LSA and CVM. Second, while the proposed policies by MABS are multidimensional and very difficult for humans to uniquely select, the coupling target function allows the preference policy proposal to be uniquely selected. Thus, this proposed approach fills a gap in the academic field of policy-making as there has never been an approach that links subjective values with objective simulation. Our approach is not limited to LSA and MABS but can be further developed to combine other subjective and objective approaches.

The remainder of this paper is structured as follows. The next section presents a literature review on subjective and objective policy evaluation approaches, followed by the Methods section that details the policy decision-making method (SOPMA) coupling subjective and objective value indices. The Results section presents the selection of policies using data obtained in Takaharu Town, Miyazaki Prefecture, Japan. This is followed by a Discussion section on the subjective value indices that should be obtained based on the results and the gap between the social issues perceived by the residents and their well-being. The final section presents the conclusions and future tasks.

## Literature review

This section provides a brief literature review of subjective and objective policy evaluation approaches. The main subjective policy evaluation approaches are the LSA [11], CVM [12], and HPM [13].

Proposed by economist Frey in 2009, LSA provides for a questionnaire, based on which a target function is established with subjective wellbeing ($f$) as the response variable and non-market goods ($x$) and income ($y$) as explanatory variables. The Marginal Willingness To Pay ($-dy/dx$) for non-market goods is then obtained from the partial derivatives ($\partial f/\partial x, \partial f/\partial y$) of $x$ and $y$ with respect to $f$. This allows non-market goods such as the natural environment and quality of life (QoL) to be valued using subjective monetary value criteria.

CVM was first proposed by environmental economics company Ciriacy-Wantrup in 1947 [19]. A widely used method in the field of environmental impact assessment and natural disaster management policies [20], CVM assesses the value of the environment as a monetary value by exploring the willingness to pay to protect the environment through a questionnaire. This allows for a comparative analysis of the costs of environmental policies and the benefits gained from the policies.

HPM was first proposed by the economist Haas in 1922 [21] and the theory was developed by Lancaster and Rosen [22] [23]. Based on the idea that the price of a good is explained by its attributes, this method collects data on the price and attributes of the good and uses statistical analysis to obtain a function of its price with the attributes as explanatory variables. This allows the estimation, for example, of the effect of the environment on the price of goods.

Subjective approaches to policy evaluation, such as LSA, CVM and HPM, allow for the evaluation of environmental policies by, for example, converting the unitary value of the environment into a monetary economic value. However, the solution of actual social issues involves not only environmental values converted into economic values, but also multifaceted economic values such as regional economic cycles, external economies, and the related population, as well as social values such as human relations, customs, and culture; consequently, it is necessary to take into account diverse and multidimensional values.

Objective policy evaluation approaches include EBPM and MABS. EBPM was facilitated by the rise of evidence-based medicine in the 1980s, and evidence-based policy has seen significant expansion since the 2000s [24]. The main processes of EBPM consist of agenda setting, which determines the priorities of social issues; policy formulation, which determines policy options based on evidence; implementation, which involves implementing action research and pilot projects; and evaluation, which involves monitoring and evaluation [25]. Evidence includes hard evidence of objective data and soft evidence of subjective data, and policymakers are expected to have a detailed and comprehensive understanding of the various policy options before making a decision on which option to choose.

MABS emerged with the mathematician Neumann's Theory of Self-Reproducing Automata published in 1966 [26], and was developed through political scientist Axelrod's prisoners dilemma model [27] and Reynolds' herd behavior model [28] in the 1980s. MABS simulates modeled autonomous agents, using their behavior and interaction rules. It allows the behavior of the whole system to be evaluated and has been applied in various fields such as transport, supply chains, consumption behavior, labor management, asset management and policy making.

Objective approaches to policy evaluation, such as EBPM and MABS, require a comprehensive understanding of a large number of policy options. However, diverse and multidimensional policy options emerge depending on the combination of hard and soft evidence weightings in EBPM and the combination of model parameters in MABS. The options chosen depend on what the policymaker considers to be the target function, and the target is multidimensional, so there is a concern that policies cannot be uniquely determined or are left to the arbitrary choice of the policymaker.

This study does not determine policies according to a unitary value converted into economic value, as in the traditional subjective policy evaluation approach, nor does it arbitrarily determine policies from a large number of policy options, as in the traditional objective policy evaluation approach. Its aim is to determine the policies that are most in line with residents' values from among the diverse and multidimensional candidate policies by setting up a new subjective–objective coupling target function combining both approaches.

# Methods

## Evaluation field

Takaharu Town in Miyazaki Prefecture was selected as the field to evaluate the proposed methodology. Located in the south-west of Japan, this town is a mountainous area, characterized by abundant water and lush greenery (colored area in Fig 1). It has a population of around 10,000 whose main livelihood is livestock farming. The residents have a long religious history and actively utilize natural resources. The population is aging very rapidly and faces challenges such as a stagnant local economy and falling birthrates.

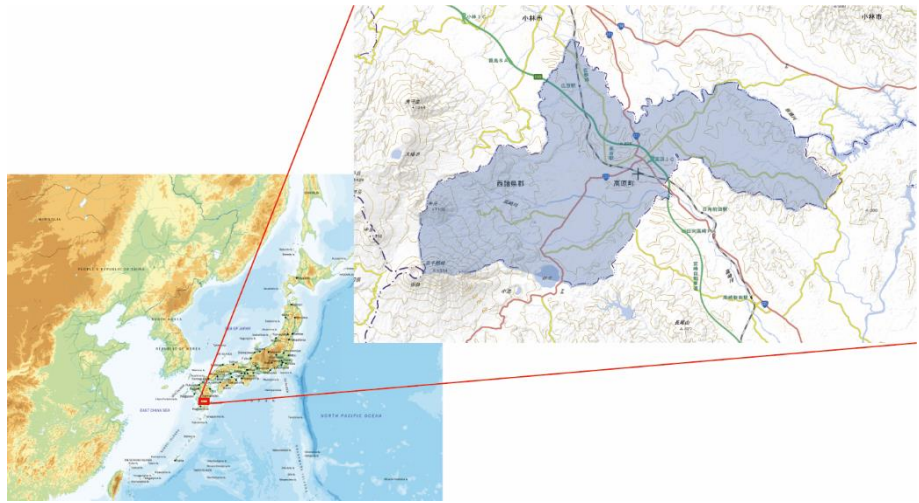

**Fig 1. Subjective questionnaire target area.** Takaharu Town is a mid-mountainous region located in the south of Japan, where water and greenery are abundant. The maps are sourced from the Geospatial Information Authority of Japan (GSI) (https://maps.gsi.go.jp/).

## Questionnaire survey

Takaharu Town conducted a "Questionnaire survey on community 'Energy

consumption' and 'Connections' in Takaharu Town. " The purpose of this survey is to investigate the actual energy use, lifestyles, and people-to-people connections of the residents of Takaharu Town. Specifically, a total of 44 questions were asked regarding the following three points.

(1) What energy consumption behavior is taking place?

(2) How are local "connections" and energy consumption related?

(3) How is the well-being of local residents and the sustainability of the community related?

This survey was distributed by posting to all households (approximately 3,900 households) living in Takaharu Town. The number of valid responses was 483 households, for a valid response rate of 11.7%. In conducting the questionnaire, we have gone through an ethical review process by Kyoto University Psychological Science Unit, and we have clearly stated that the questionnaire responses are left to the free will of the respondents themselves.

## Subjective target function

First, consider the representation of the subjective values of the residents on the basis of their responses to several questionnaires. Specifically, suppose that residents gave the answer $x_i$ $(i = 1, 2, ..., n)$ to $n$ questions $q_i$ $(i = 1, 2, ..., n)$ in the questionnaire. The subjective values of the residents, the subjective target function $y$, are then expressed in equation (1).

$$y = f(x_1, x_2, x_3, \cdots, x_i, \cdots, x_n) \qquad (1)$$

The form of the function in equation (1) can take many forms, but assume that it can be expressed by a multiple regression equation, with $x_i$ as the explanatory variable and where, as explanatory variables in the multiple regression equation, responses that are highly correlated with the subjective response variable are used.

$$r = \frac{\sum_{i=1}^{n}(x_i - \bar{x})(y_i - \bar{y})}{\sqrt{\sum_{i=1}^{n}(x_i - \bar{x})^2}\sqrt{\sum_{i=1}^{n}(y_i - \bar{y})^2}} \qquad (2)$$

A correlation analysis of the results of the awareness survey questionnaire was conducted using the results of the question: "How well-being are you at present?" Question items with high correlation coefficients to the results of responses to the question item (Questionnaire on subjective well-being) was extracted.

Based on the results of the multiple regression analysis, the target function $y$ in equation (1) can be expressed in a linear regression equation as equation (3). From the correlation coefficient $r$ obtained using equation (2), the linear regression equation is expressed using only those responses that are highly correlated with the target function $y$. From these, the subjective response function $y$, which reflects the results of the questionnaire, can be expressed as in equation (3) using the explanatory variable $x_i$ based on the responses. where $\beta$ is the partial regression coefficient of the multiple regression equation.

$$y = \beta_0 + \beta_1 x_1 + \beta_2 x_2 + \cdots + \beta_i x_i + \cdots + \beta_n x_n + \varepsilon \qquad (3)$$

From these questions, two items each related to humans, society, ecology and economy were extracted. The questions relating to humans were related to the individual, those relating to society were related to the area and residents, those relating to ecology, to the natural environment and those relating to economy, to work and household finances. The multiple regression equation for the target function $y$ with these eight explanatory variables is equation (4).

$$y = \beta_0 + \beta_1 x_1 + \beta_2 x_2 + \beta_3 x_3 + \beta_4 x_4 + \beta_5 x_5 + \beta_6 x_6 + \beta_7 x_7 + \beta_8 x_8 \qquad (4)$$

## Objective policy simulation

Next, we will explain the method of policy simulation, an objective evaluation approach. MABS, which uses the interaction of multiple agents as shown in Fig 2, is used to simulate the policies. Local residents, local companies, the natural environment, among other are set as agents. Each agent acts according to its own rules and attribute information, with interactions between agents occurring at specific times. This generates a complex intertwining behavior of the individual elements of the system as a whole. For a given policy, given the agents' rules and attribute information as parameters, evaluation indices reflecting the overall behavior of the system can be calculated.

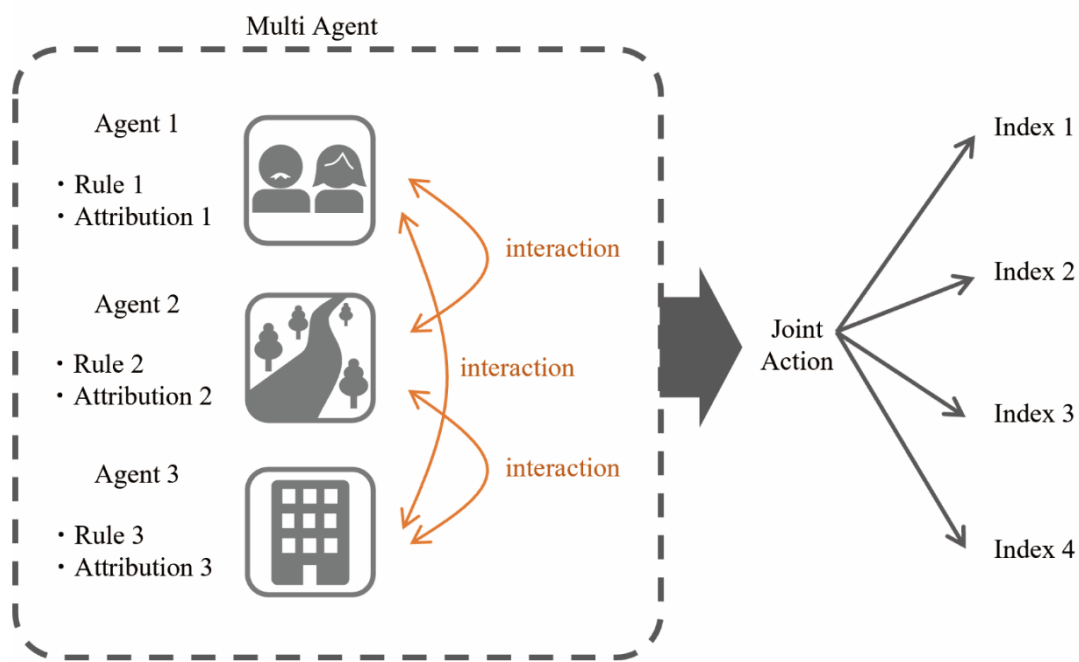

**Fig 2. Conceptual diagram of a multi-agent-based simulation (MABS).** Agents 1, 2, and 3 have different rules and attribute information. The interaction of each agent produces the behavior of the whole system, for which multiple and pluralistic

evaluation indices are calculated.

Specifically, this study will target policies to enhance regional sustainability through energy local production for local consumption. A simulation using MABS will be used to determine proposed policies for local production and consumption of energy using natural energy for the Takaharu Town, which is rich in natural resources. Fig 3 presents the simulation model that assumes local production and local consumption of energy from renewables. The model uses as sources natural energy—nano-hydropower and photovoltaics—in the area, plus the use of storage batteries and, if necessary, electricity supply from a major power company.

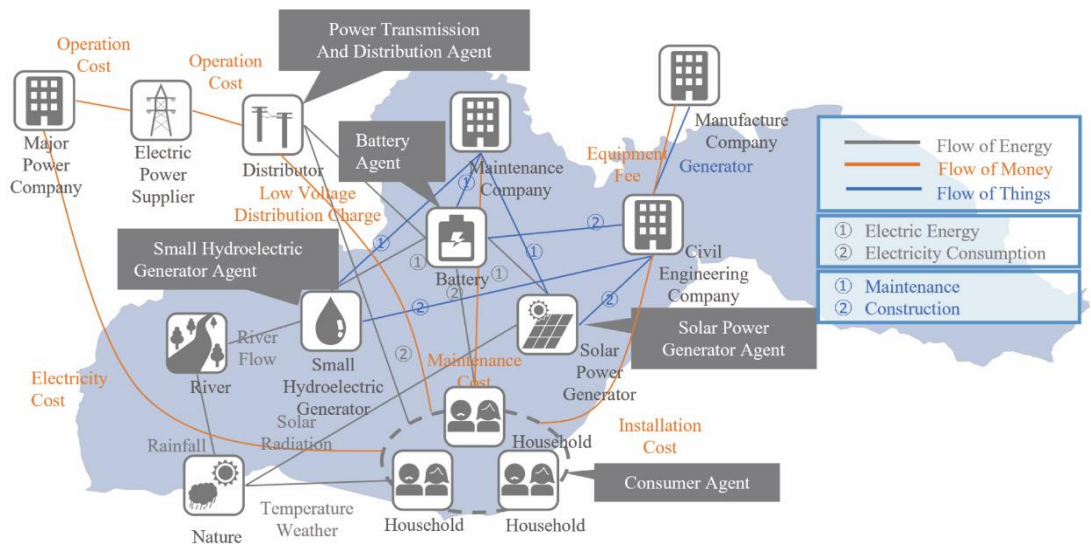

**Fig 3. Model for local production and local consumption of energy.** Each icon in the figure represents an agent, for example, the solar power generator, small hydroelectric generator, battery, power transmission, and consumer agents. Money, things, and energy are exchanged between agents.

To simulate the proposed policies, several pyranometers were installed in Takaharu Town, as shown in Fig 4, to measure solar radiation and calculate the potential for solar power generation. Water level meters were installed to measure river levels and flow rates and calculate the potential for nano-hydropower generation. Data from these sensors were collected for about a year and used to calculate the potential for renewable energy generation in Takaharu Town.

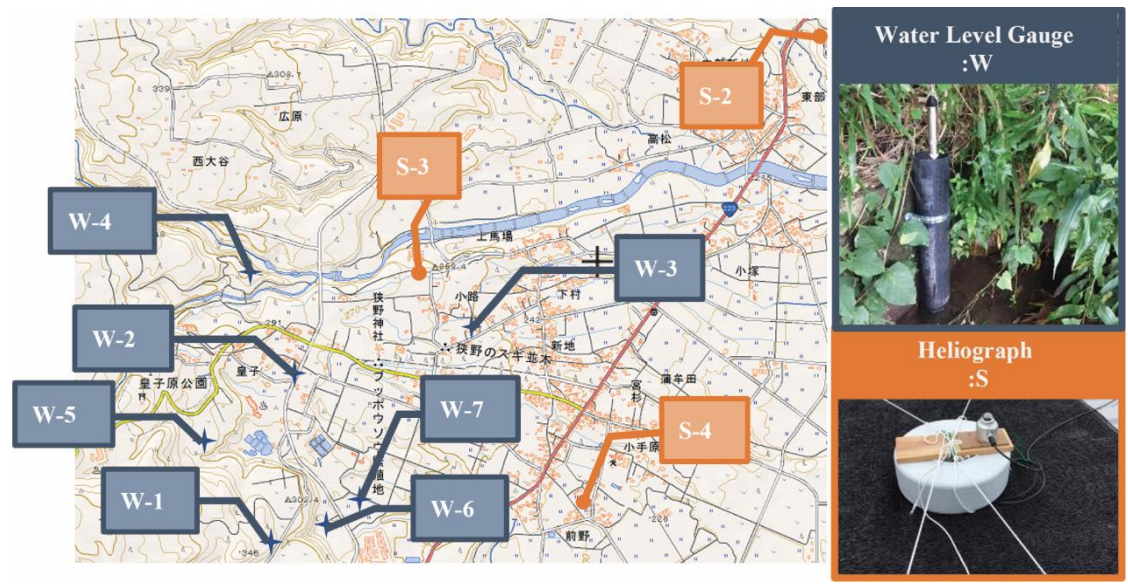

**Fig 4. The locations and numbers of water level and solar radiation gauges or sensors installed in Takaharu town.** W is the water level gauge and S is the solar radiation gauge. The maps are sourced from the GSI (https://maps.gsi.go.jp/).

Furthermore, the obtained potential for generating renewable energy was used to simulate policies for introducing power generation facilities in the local production and consumption model, in which the energy generated is consumed locally. The scale of the power generation facilities, such as photovoltaic, nano-hydro, and storage batteries to be installed, was given as a variable, and the amount of renewable energy generation was calculated at hourly intervals over a one-year period. The amount of energy generated by renewables was then compared with the amount of energy used by residents, and evaluation indices such as the utilization rate of renewables, the cost borne by residents (electricity bills) and the rate of economic circulation within the region as a result of the introduction of power generation facilities were calculated.

Next, we provide an overview of the regional policy simulation to calculate potential policies for the introduction of renewable energy generation facilities in the region. Multiple agents are set up as follows, and while exchanging data between agents, multiple candidate policies and their evaluation indices are considered as output.

- Photovoltaic models: generator size, generation (forecast), cost and service life.
- Hydropower models: generation size, generation (forecast), cost and service life.
- Rechargeable battery models: capacity, cost, service life.
- Electricity bills, electricity transport charges (low voltage consignment charges).
- Air temperature, room temperature
- Electricity use model: power consumption of air conditioners relative to room temperature and temperature setting.
- Electricity demand model: air conditioner temperature set by the user.

The above agents are used to output the intra-regional distribution rate of the local economy, the costs borne by the residents, and the utilization rate of renewable energy as evaluation indices for candidate policies. Fig 5 illustrates the data flow diagram of the regional policy simulator. First, a cleansing process is carried out by deleting unnecessary and missing data from the sensor data. In case of missing data, data from other similar regions is used. Next, the cleansed data is used to generate a predictive model of electricity generation and energy use in the region. Finally, simulations are carried out under different conditions, such as the scale of power generation by generators, to calculate the evaluation indices.

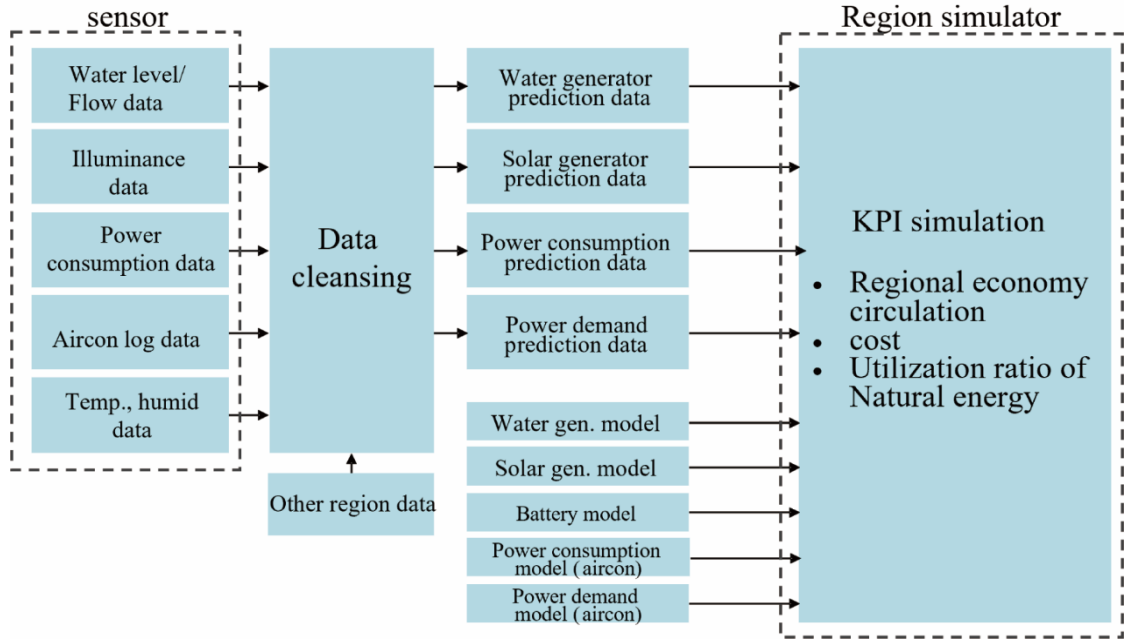

**Fig 5. Block diagram of data flow for objective policy simulation.** Cleansing sensor data and creating models. Using the created model, simulations will be conducted to calculate KPIs such as regional economic circulation, costs, and natural energy utilization rates.

The set of parameters $P_k$ as input to this policy simulator and the set of evaluation indices $A_k$ of the policies obtained by the simulation can be expressed in equation (5) where, $k$ is the number of candidate policies obtained by simulation, $\mu$ is the number of input parameters, $m$ is the number of indices to be calculated and $l$ is the total number of candidates to be output.

$$P_k = \{(p_{1k}, p_{2k}, \cdots, p_{jk}, \cdots, p_{\mu k}) | k = 1,2, \cdots, l\}$$
$$A_k = \{(a_{1k}, a_{2k}, \cdots, a_{jk}, \cdots, a_{mk}) | k = 1,2, \cdots, l\} \quad (5)$$

## Subjective–objective coupling target function

Finally, we explain our proposed method of narrowing down candidate policies by coupling objective value indices with subjective values. $\chi_{ik}$ is obtained by coupling $A_k$ with $x_i$, as shown in equation (6), where $A_k$ is the set of objective value indices calculated by the policy simulation in equation (5) and $x_i$ is the explanatory variable for the target function $y$ obtained by the questionnaire in equation (1). Suppose that zero or more objective value indices $a_{jk}$ interact for some explanatory function $x_i$.

$$\chi_{ik} = g(x_i, a_{jk}) \tag{6}$$

For the target function $y$ with the subjective value index as the explanatory variables $x_i$, the subject–object coupling target function $\psi_k$ with the explanatory variables $\chi_{ik}$ incorporating the objective value indices are expressed in equation (7). If the subjective target function $y$ is expressed in linear form as in equation (3), the coupling target function $\psi_k$ is expressed in equation (8).

$$\psi_k = \psi(\chi_{1k}, \chi_{2k}, \cdots, \chi_{ik}, \cdots, \chi_{nk}) \tag{7}$$

$$\psi_k = \beta_0 + \beta_1 \chi_{1k} + \beta_2 \chi_{2k} + \cdots + \beta_i \chi_{ik} + \cdots + \beta_n \chi_{nk} + \varepsilon \tag{8}$$

As shown in equation (9), the policies can be narrowed down from the many candidates shown in equation (5) by finding the policy $k_{opt}$ that optimizes this coupling target function $\psi_k$. Furthermore, the parameter set $P_{k_{opt}}$ and evaluation index set $A_{k_{opt}}$ for the selected policy $k_{opt}$ are obtained as in equation (10).

$$k_{opt} = \underset{k}{\mathrm{argmax}}\, \psi_k \qquad (9)$$

$$P_{k_{opt}} = \left\{ \left( p_{1k_{opt}}, p_{2k_{opt}}, \cdots, p_{jk_{opt}}, \cdots, p_{\mu k_{opt}} \right) \right\}$$

$$A_{k_{opt}} = \left\{ \left( a_{1k_{opt}}, a_{2k_{opt}}, \cdots, a_{jk_{opt}}, \cdots, a_{mk_{opt}} \right) \right\} \qquad (10)$$

Specifically, the following three hypothetical resident value types are established to couple the economic circulation rate in the region (social value), the natural energy utilization rate (ecological value), and cost borne by residents (economic value), which are the evaluation indices of the policy simulation, into the linear regression equation obtained in equation (4).

- Type A : Rising electricity prices decrease economic wealth.
- Type B : While the use of renewables increases pride and increases in the local economic circulation rate enriches the economy, higher electricity prices decrease economic wealth.
- Type C : Rising electricity prices reduce economic affluence, and the use of natural energy undermines attachment to the district. On the other hand, through the use of natural energy and local economic circulation, the trust of district residents is increased.

Substituting these types A to C as interactions for the explanatory variables in equation (4), we obtain the subject-object coupling target function $\psi_k$. These subjective-objective coupling target functions, $\psi_k$, are the optimal policies for each type of value by calculating the policy with the highest subjective well-being according to equation (9).

# Results

## Subjective target function

The result of the "How well-being are you at present?" question item in the residents' questionnaire were used as the response variable, and correlation analysis was conducted with the results of the other questions. The correlation coefficient $r$ was calculated using equation (2), we extracted the question items for which have a positive correlation of $r = 0.1$ or more and a significance level of $p \leq 0.05$. Table 1 shows a selection of two questions related to "human," "society," "ecology," and "economy" from each of these categories. Of the 483 resident questionnaires collected, we used data from 421 questionnaires that had complete data for these eight questions.

**Table 1. List of Extracted Questionnaires and Classifications**

| Questionnaire item | Classification |
|---|---|
| How is your current health status? | Human 1 |
| I trust people who live in the same district. | Society 1 |
| I have an attachment to this district. | Society 2 |
| I think I am making my loved ones happy. | Human 2 |
| Environmentally friendly behavior is socially required. | Ecology 1 |
| Acting in an environmentally friendly manner is a source of pride. | Ecology 2 |
| Where is your place of work/office located? -In the district | Economy 1 |
| Compared to others in your district, how financially affluent are you? | Economy2 |

Table 2 and Fig 6 show the results of multiple regression analysis using a linear regression model for the subjective well-being question items, with the maximum value of each question item normalized to be 1 for the 421 questionnaire data with 8 questions. From the results shown in Table 2, equation (11) is expressed using a linear regression equation for the subjective target function when subjective well-being is the response variable. Furthermore, the horizontal axis in Fig 6 is the subjective well-being obtained from the questionnaire, normalized in the range 0-1, and the vertical axis is the predicted value calculated using the linear regression equation presented in equation (11). The coefficient of determination of the regression equation using the questionnaire data is $R^2 \fallingdotseq 0.37$ (equivalent to ~0.6 in terms of correlation coefficient), which is high enough in the field of policymaking approaches such as LSA and the proposed method [29] [30].

**Table 2. Linear Regression Analysis Results**

| Item | Coefficient(β) | Standard Error | t | P-value |
|---|---|---|---|---|
| Intercept | 0.12842 | 0.055365 | 2.31953 | 0.020854 |
| Human 1 (x1) | 0.394934 | 0.036739 | 10.7497 | 6.34E-24 |
| Human 2 (x2) | 0.158204 | 0.048217 | 3.28111 | 0.001122 |
| Society 1 (x3) | 0.107405 | 0.053776 | 1.997277 | 0.046452 |
| Society 2 (x4) | -0.02598 | 0.04611 | -0.56343 | 0.573447 |
| Ecology 1 (x5) | 0.003615 | 0.056051 | 0.064494 | 0.948608 |
| Ecology 2 (x6) | -0.00477 | 0.046172 | -0.10335 | 0.917738 |
| Economy 1 (x7) | 0.048433 | 0.018388 | 2.63404 | 0.008755 |
| Economy 2 (x8) | 0.252739 | 0.052488 | 4.815139 | 2.07E-06 |

$$y = 0.12842 + 0.394934 \cdot x_1 + 0.158204 \cdot x_2 + 0.107405 \cdot x_3$$
$$- 0.02598 \cdot x_4 + 0.003615 \cdot x_5 - 0.00477 \cdot x_6 \qquad (11)$$
$$+ 0.048433 \cdot x_7 + 0.252739 \cdot x_8$$

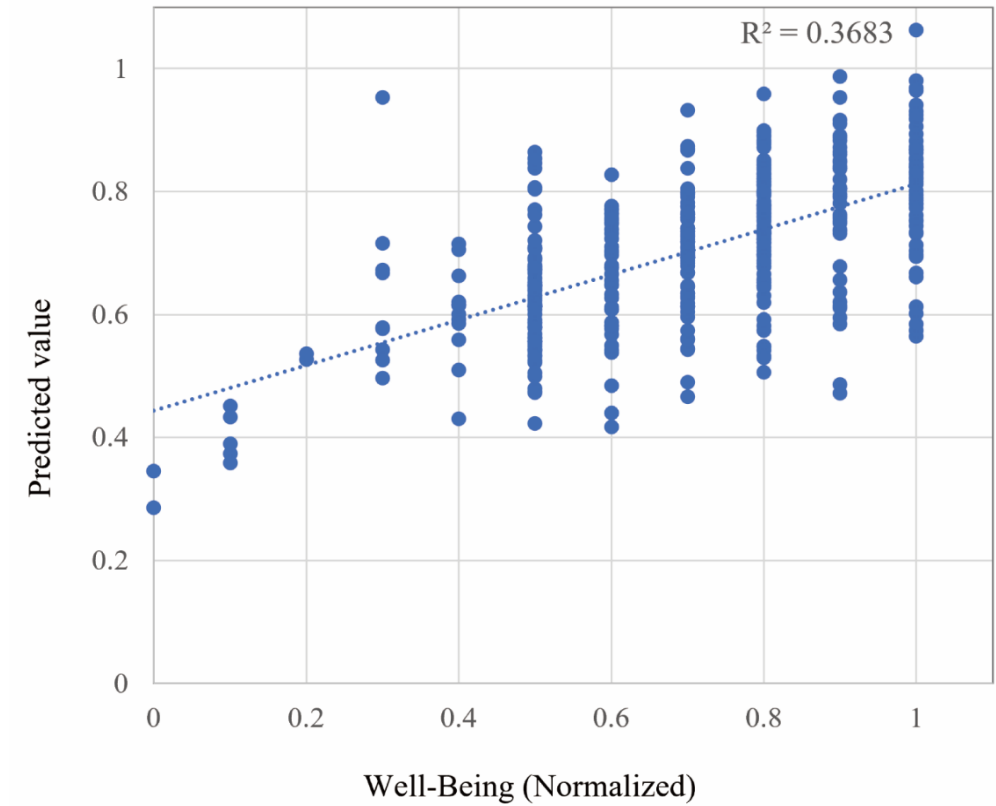

**Fig 6. Linear regression analysis results and subjective target function.** The horizontal axis is the value of responses to "subjective well-being" questions normalized from 0 to 1. The vertical axis is the predicted value of the subjective target function calculated using the linear regression equation shown in equation (11). The plots are the 421 data obtained from the questionnaire.

## Objective policy simulation

Twenty thousand proposed policies were created under changing combinations of the scale of photovoltaic power generation facilities, the scale of nano-hydro power generation facilities and the capacity of storage batteries as conditions for the implementation of local production and consumption of renewable energy in the local region. The policy evaluation indices rate of economic circulation in the region, rate of use of renewable energy and cost borne by the residents were defined as social value, ecological value and economic value, respectively. To evaluate which value axis each of the 20,000 simulation results leaned toward, they are shown in the triangular graph in Fig 7. Each plot on the triangular graph is a policy obtained from the simulation. The tops of the triangular graph represent social value, ecological value and economic value, respectively, and the plots closest to each top are policies that focus on each of these values. In other words, plots close to the top of the economic value are policies that focus on economic value rather than social or ecological value. Therefore, a plot close to the center of gravity of the triangle is a balanced policy of the three values. However, as the triangular graph is designed to express the ratio of social, ecological and economic values of a policy, it cannot express differences in the magnitude of the values of each value indices.

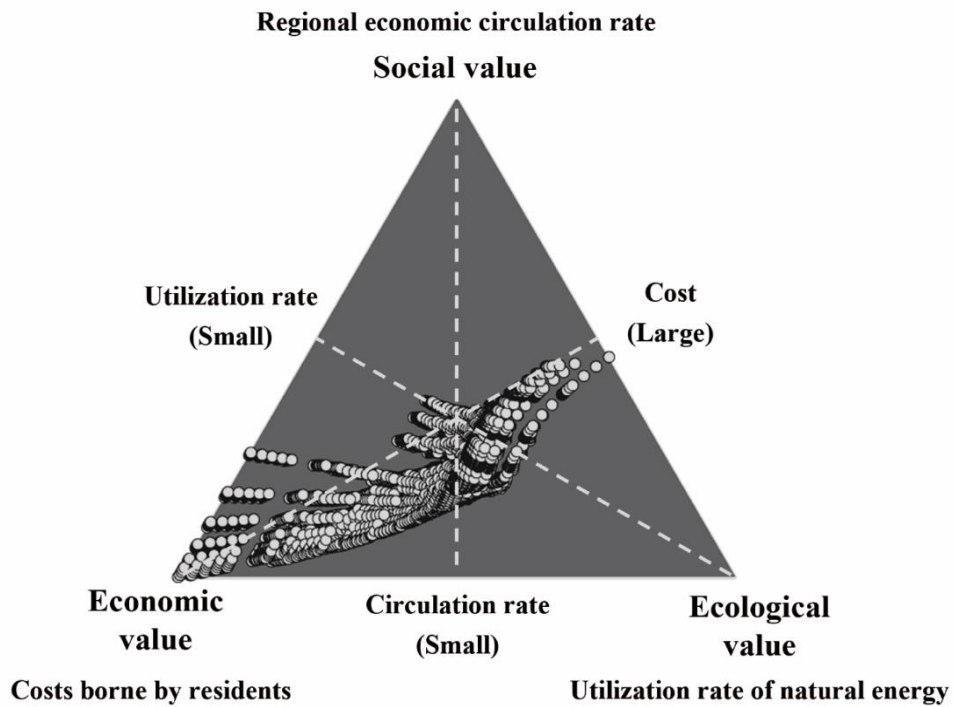

**Fig 7. Three-value evaluation of objective policy simulation results.** The triangle graph represents the balance of the three value axes. Therefore, a plot closer to each top is a policy that places more emphasis on that value axis. Overall, the simulation results show that many of the policies are weight toward economic value.

## Subjective–objective coupling target function

For the policy simulation results, a target function on subjective well-being obtained from the questionnaire was used to predict the impact of the introduction of the policy on residents' well-being. In order to couple the evaluation index of the simulation to the subjective target function, assuming residents' value types A to C and substituting the evaluation index as an interaction for each type of explanatory variable, the subjective-objective coupling target function $\psi_k$ becomes equations (12)–(14). These equations correspond to equation (7) or equation (8) in the Methods section, where the value $x_{iav}$ for each question item on the right-hand side was the mean value of the 421 data. The share of household expenses on electricity in Japan is 0.037 (2019 data) [31], and the non-electricity household expense was calculated as 0.963. The $p$ used in each equation represents the cost borne by the residents, $u$ represents the renewables utilization rate, and $d$ represents the intra-regional distribution rate of the local economy. In this case, $p_0$, $u_0$ and $d_0$ represent the current situation without the use of renewables, before the introduction of the policies.

$$\psi_A = 0.12842 + 0.394934 \cdot x_{1av} + 0.158204 \cdot x_{2av} + 0.107405$$
$$\cdot x_{3av} - 0.0259796 \cdot x_{4av} + 0.00361494 \cdot x_{5av}$$
$$- 0.00477173 \cdot x_{6av} + 0.0484335 \cdot x_{7av} \qquad (12)$$
$$+ 0.252739 \cdot x_{8av} \cdot \frac{1}{0.963 + 0.037 \cdot \frac{p}{p_o}}$$

$$\psi_B = 0.12842 + 0.394934 \cdot x_{1av} + 0.158204 \cdot x_{2av} + 0.107405$$
$$\cdot x_{3av} - 0.0259796 \cdot x_{4av} + 0.00361494 \cdot x_{5av}$$
$$- 0.00477173 \cdot x_{6av} \cdot \frac{1+u}{1+u_o} + 0.0484335 \cdot x_{7av} \qquad (13)$$
$$+ 0.252739 x_{8av} \cdot \frac{1+d}{1+d_o} \cdot \frac{1}{0.963 + 0.037 \cdot \frac{p}{p_o}}$$

$$\psi_C = 0.12842 + 0.394934 \cdot x_{1av} + 0.158204 \cdot x_{2av} + 0.107405$$

$$\cdot x_{3av} \cdot \frac{1+d}{1+d_0} \cdot \frac{1+u}{1+u_0} - 0.0259796 \cdot x_{4av} \cdot \frac{1+u}{1+u_0}$$

$$+ 0.00361494 \cdot x_{5av} - 0.00477173 \cdot x_{6av} \cdot \frac{1+u}{1+u_o} \quad (14)$$

$$+ 0.0484335 \cdot x_{7av} + 0.252739 x_{8av} \cdot \frac{1+d}{1+d_o}$$

$$\cdot \frac{1}{0.963 + 0.037 \cdot \frac{p}{p_o}}$$

Using these subjective–objective coupling target functions $\psi_k$, equations (12)–(14), the policies with the highest subjective well-being was calculated according to equation (9). Fig 8 shows these results overlaid on the three value assessment results of the policy simulation results presented in Fig 7. Table 3 presents the subjective–objective coupling response variables (subjective well-being) obtained using the regression equations presented in equations (12)-(14), and the input parameters and output evaluation indices of the simulations for type A, B and C, respectively. These results correspond to equation (10) in the Methods section.

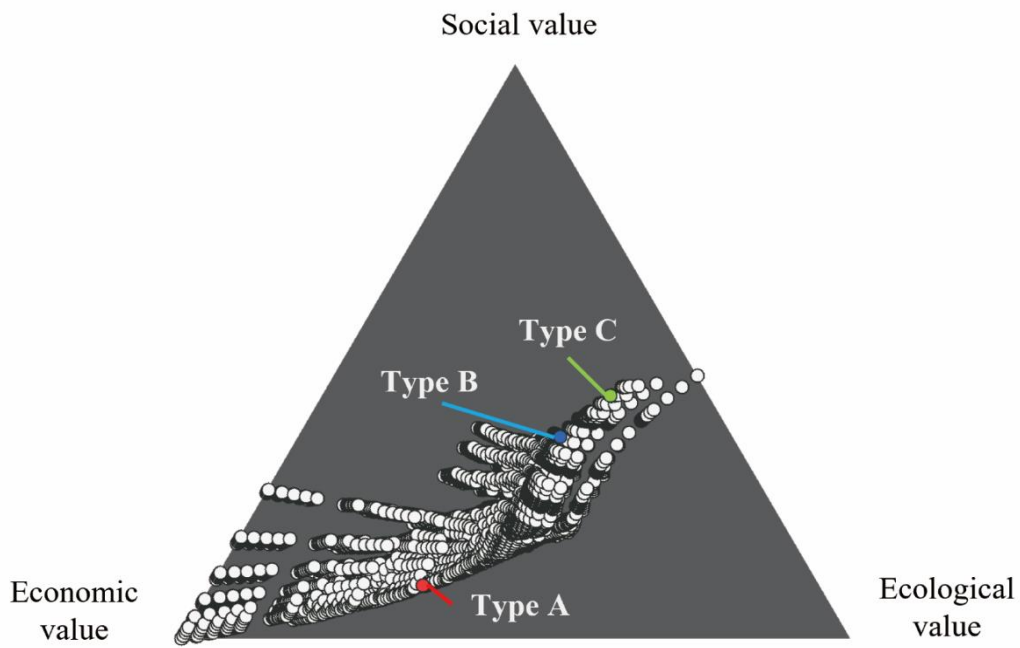

**Fig 8. Optimal policy for different types of subjective–objective coupling target functions.** This is the result of the selection of optimal policy proposal for each of the three hypotheses A, B and C. In terms of the balance of the three values, Type A has low social value and focuses more on ecological and economic values. Type B has higher social and ecological value than type A, but lower economic value. Type C has the same ecological value as type B but higher social and lower economic value.

**Table 3. Subjective–Objective Coupling Variables, Simulation Parameters and Results.**

| Type | Now | A | B | C |
|---|---|---|---|---|
| Subjective–Objective Coupling Variable ($\psi$) | 0.702375 | 0.703907 | 0.768115 | 0.912414 |
| Cost [¥] | 4,157,930 | 2,908,197 | 10,183,154 | 16,978,103 |
| Regional Distribution Ratio | 0 | 10.3 | 58.9 | 64.8 |
| [%] | | | | |
| Renewable Energy Utilization Rate [%] | 0 | 53.0 | 99.5 | 100 |
| Annual Power Generation (Hydro) [kWh]/ Effective Dropout [m] | 0/ 0 | 58,177.3/ 2 | 581,773/ 20 | 2,327,091/ 80 |
| Annual Power Generation (Solar) [kWh]/ Generation Capacity [kW] | 0/ 0 | 21,029.06/ 20 | 841162.6/ 800 | 630871.9/ 600 |
| Storage Battery Capacity [kWh] | 0 | 0 | 0 | 0 |
| Electricity Purchased from the Power Company [¥] | 4,157,930 | 1,912,632 | 886,561 | 874,800 |

As Hypothesis A assumes that higher electricity bills reduce economic wealth, the policy with the lowest cost to residents among the candidate policies was selected as the policy that would increase residents' well-being. The policy was to install small-scale hydroelectric and photovoltaic installations and to purchase electricity from major power companies for about half of the annual electricity requirements. This result appears to indicate a temporary increase in the cost of installing new equipment. However, in the nature-rich Takaharu Town, even a small-scale power generation facility could cover half of its annual electricity consumption with renewables. Moreover, when electricity prices and installation costs are considered together, it would be cheaper than current electricity use relying solely on the power company. This policy is also expected to result in a small but significant economic redistribution in the region because of the use of renewables.

Next, Hypothesis B chooses a policy that increases the rate of economic operations in the region with a high rate of use of renewables, while at the same time limiting a significant increase in electricity bills. The policy is to increase the utilization rate of natural energy by introducing the largest photovoltaic installations and some hydropower installations, thereby increasing the rate of economic circulation in the region, while reducing the costs borne by the residents. Therefore, they do not rely on renewables for all the electricity used in the region, but purchase a certain amount of deficient electricity from the major power companies.

Finally, Hypothesis C is a policy that increases the electricity costs compared to Hypothesis B, but also improves the rate of economic circulation in the region, with a 100% utilization rate of renewables by introducing hydropower generation.

There is a major difference between Hypothesis C and Hypothesis B, with the former introducing maximum hydropower installations and the latter maximum photovoltaic installations. Hydroelectric installations are more effective than photovoltaic installations in increasing the rate of economic circulation in the region but have the effect of increasing costs to residents. In other words, under hypothesis C, promoting the use of renewables and local revitalization, even at an economic cost, is a policy that increases the subjective well-being of the residents.

## Discussion

By coupling simulation methods using objective data, which have conventionally been used for policy evaluation, with subjective value evaluation methods using questionnaires in social psychology, etc., this study showed SOPMA can be used to select preferred policies, taking both subjectivity and objectivity into account.

As for conventional policy evaluation methods, the simulation method used here to calculate the complex behavior produced by the interaction between multiple agents using MABS was used to calculate approximately 20,000 policies proposals for the use of renewable energy. While simulations can calculate a large number of proposed policies, there are so many that it is very difficult to narrow down the best policies for the region. In the questionnaire, the results of the question items related to Human, Society, Ecology, and Economy were extracted from the results of other question items that were highly correlated with the results of the subjective well-being question items, and a multiple regression equation with subjective well-being as the response variable was presented as the subjective target function. For the eight explanatory variables of the derived subjective target function, the values of the evaluation indices obtained in the simulation are coupled. By coupling simulation technology, an engineering approach, with questionnaire analysis, a social psychological approach, it showed the feasibility of taking into account residents' preferences in selecting regional policies that have a large impact on them.

The renewables energy models used in this simulations only incorporate specific power generation models such as nano-hydro and photovoltaics. Since different regions have different natural environment, other renewable energy generation, such as biomass and wind power would also need to be considered in the future. Furthermore, three indices were used as the main evaluation indices for the simulation: the cost to residents, the utilization rate of renewable energy, and the local economic circulation rate, but there are other possible indices to be considered, such as $CO_2$ emissions and the intra-regional employment rate. This means that the objective simulation model can be further refined according to the goals of the policy and the region.

In addition, subjective well-being was set as the response variable for the questionnaire in this study, it is assumed that the required response and explanatory variables will differ depending on the target area and the goals of the policy. In this study, three value types of assumed residents, A, B, and C, were examined. The results showed that policies to increase subjective well-being differed for each type. In the actual community, there are more stakeholders with different generations, professions, positions, etc., and with different values. In order to determine policies as a region and introduce then to the region, it is necessary not only to understand that there are different values, but also to consider how they should be determined as a region. In addition, in order to couple objective and subjective data more smoothly, it is expected the coefficient of determination of the subjective target function and the accuracy of the subjective-objective coupling target function can be improved by creating questionnaire items that assume the combination of the objective and subjective approaches in advance.

Other methods than multiple regression analysis of subjective value indices based on questionnaire results could be used to obtain subjective values. One method is the logic model, a social impact evaluation method that reflects subjective values. This combination of various approaches makes it possible to evaluate policies that couple the subjective and objective. This will lead to further developments regarding the coupling of not only subjective LSA and CVM with objective EBPM and MABS but also other subjective and objective approaches.

Finally, in actual communities, regions, and companies, where stakeholders with different values and positions coexist, multiple policies can be selected if SOPMA is conducted based on their respective values. After each stakeholder has narrowed down the optimal policy options by using SOPMA, a better social choice can be made by creating a consensus-building forum for constructive discussions based on the policy options.

## Conclusions

This study presents a new subjective–objective policy decision-making approach, SOPMA, with the goal of proposing a method for selecting the most appropriate policy proposal, taking into account the subjective values of residents, for a large number of local policy proposals calculated using big data, such as sensor and open data.

As a real field to evaluate this approach, Takaharu Town in Miyazaki Prefecture, Japan, was selected to test a renewable energy introduction policy. As a method for extracting the residents' subjective values, a subjective target function was established from an awareness survey questionnaire conducted among the residents, with the response variable being the answers relating to subjective well-being and the explanatory variables being the other answers highly correlated with this response variable. Further, using objective sensor data, a policy simulation using MABS was carried out to calculate approximately 20,000 policy proposals. The results showed that in areas such as Takaharu Town, where people take pride in their local environment, there is a high preference for the use of renewable energy in a way that does not damage nature, which means balancing the costs borne by the residents with the rate of renewable energy utilization would have a significant impact on policy decisions. Further, stimulating the local economy through the use of renewables improves the subjective well-being of the residents, compared to the increase in the costs.

We found that SOPMA can be used to extract policy proposals reflecting local culture, characteristics and attitudes from among a large number of policy proposals by defining a subjective–objective coupling target function that takes into account the interaction between the subjective target function obtained from the results of the local awareness survey questionnaire and the objective evaluation indices obtained from policy simulation.

In the past, it was very difficult to ascertain which policies were appropriate, even though the various stakeholders supporting the region had different opinions and could assess the merits and demerits of a particular proposal. However, the new method, SOPMA, makes it possible to concretely express policy proposals reflecting the values of each stakeholder, and promote more constructive discussions for consensus-building.

In this study, the definition of a subjective–objective coupling target function, which combines a subjective target function and an objective evaluation index, gives rise to the possibility of selecting policy proposals that take subjective values into account. However, different generations, occupations, and positions may change the question items to highly correlate with their subjective well-being, so that the content and direction of the policies to be implemented will differ significantly. Therefore, future research is required to examine methods and items, such as questionnaires, to obtain subjective values to narrow down proposed policies with high precision, as well as the ensure validity of the objective policy evaluation indices.

Further, by using the subjective–objective coupling target function of SOPMA, it may be possible to extract concrete policy proposals that reflect the subjective values of the stakeholders in the region. However, to select one policy that is necessary and implementable for a region, a consensus-building forum is needed to ensure that the policy is acceptable to each stakeholder. It is important to work with consensus-building support technologies that take into account the order of people's preferences, tolerance ranges, etc., and narrow them down most appropriately.

## Acknowledgements

The author would like to thank Professor Yoshinori Hiroi of Kyoto University for his great help in acquiring sensor data in Takaharu Town and Professor Yukiko Uchida of Kyoto University for the questionnaire on residents' awareness. The author is grateful to Takaharu Town Hall and the residents of Takaharu Town for their cooperation in obtaining this data and administering the questionnaire, and to the members of the Hitachi Kyoto University Laboratory for their advice and support during the writing of this paper.